# Modelling the Index of Sustainable Economic Welfare (ISEW) and its response to policies

Luzie Dallinger, Reo Van Eynde, Jefim Vogel, Lorenzo Di Domenico, Seán Fearon, Tina Beigi, Cédric Crofils, Kevin J. Dillman, and Daniel W. O'Neill

**Abstract**

Given the challenge of achieving societal welfare in an environmentally sustainable way, the Index of Sustainable Economic Welfare (ISEW) has emerged as an alternative progress indicator in response to critiques of Gross Domestic Product (GDP). The ISEW compares the benefits of economic activity with its social and environmental costs. Previous studies have analysed the past performance of the ISEW, but none have simulated the ISEW using a dynamic macroeconomic model. We address this important gap by incorporating the ISEW into COMPASS, an ecological macroeconomic model. We analyse how the ISEW is affected by three social and environmental policies (a carbon tax, income redistribution, and working-time reduction) and explore whether the ISEW is well-suited for the evaluation of these policies. We find that the ISEW grows over time in all scenarios. The strongest improvement over business-as-usual occurs when all policies are combined, with the individual policies mostly increasing the ISEW. However, in the case of working-time reduction, the ISEW decreases compared to business-as-usual, which is an unexpected finding. Our results suggest that the ISEW is better than GDP at capturing social and environmental effects, but that it underestimates the environmental costs of economic activity. We argue that a strong sustainability approach, with a comprehensive representation of biophysical boundaries and social thresholds, would provide better guidance to policymakers who aim to achieve sustainable and inclusive wellbeing.

## 1. Introduction

Over the past fifty years, the size of the global economy — usually measured by Gross Domestic Product (GDP) or GDP per capita — has expanded markedly, with real GDP per capita more than doubling (World Bank, 2025). Yet, the state of the planet has deteriorated over the same period, and living conditions remain inadequate for many (Fanning and Raworth, 2025; McNeill and Engelke, 2016; O'Neill et al., 2018). This mismatch between the main indicator used in policymaking and the realities of environmental and social wellbeing has prompted calls for alternatives to GDP since its creation (Costanza et al., 2025, 2009; Van den Bergh, 2009).

Several alternative indicators have been created in response, such as the Human Development Index (UNDP, 1990), the ecological footprint (Wackernagel and Rees, 1996), the Sustainable Development Goals (United Nations, 2015), and the "Doughnut" of social and planetary boundaries (Fanning and Raworth, 2025; Raworth, 2017). One of the more prominent alternatives is the Index of Sustainable Economic Welfare (ISEW), developed by Daly and Cobb (1989). The ISEW calculates the balance between monetary estimates of the benefits from economic activity with monetary estimates of its social and environmental impacts. It includes many elements present in GDP, such as household and government consumption, but adds further benefits, particularly from the informal sector. It also deducts penalties for unwanted consequences of economic activity, such as environmental degradation, inequality, and harmful consumption (Daly and Cobb, 1994; Soupart and Bleys, 2024).

Previous research on the ISEW has concentrated on methodological refinements (Lawn, 2003; Van der Slycken and Bleys, 2020), calculations for different geographical areas (Jackson et al., 2008; Soupart and Bleys, 2024), and analyses of past developments (Kalimeris et al., 2020; Tsara et al., 2024). However, there is a lack of forward-looking assessments of how the ISEW responds to policy changes or other societal developments. To our knowledge, our study is the first to dynamically model the ISEW and its response to policies using an ecological macroeconomic model.

We address two research questions: (1) How do three social and environmental policies — a carbon tax, income redistribution, and working-time reduction — affect the ISEW? (2) Is the ISEW well-suited for the evaluation of social and environmental policies?

The remainder of this article proceeds as follows. Section 2 reviews the literature on the ISEW and outlines the knowledge gap that this study addresses. Section 3 explains the methods used to model the ISEW and the three policies assessed. Section 4 presents the results, showing how different policy combinations affect the ISEW and comparing these outcomes with those for GDP. It also compares the results to selected social and environmental indicators. Section 5 discusses limitations of the study, the value of dynamically modelling the ISEW, and its suitability as an indicator for the evaluation of social and environmental policies. Section 6 concludes.

## 2. Literature review

We start by outlining the ISEW's historical development and major areas of application (Section 2.1). We then discuss key criticisms of the ISEW and previous efforts to address them through refinements of its theoretical foundation (Section 2.2). Next, we introduce core concepts relevant to ecological macroeconomic modelling (Section 2.3). Subsequently, we provide a broad overview of the literature on the three policies analysed in this study (Section 2.4). The final section identifies an important research gap with respect to modelling the ISEW in policy scenarios (Section 2.5).

### 2.1 ISEW development and applications

The initial idea for the ISEW stems from Daly and Cobb (1989), who observed that heavy reliance on Gross National Product (GNP) for policy assessment overlooks the negative consequences of economic growth. GNP is closely related to GDP, but measures the value of goods and services produced by factors of production owned by a country's residents, whereas GDP measures production within national borders (Argandoña, 2016). The ISEW was proposed as a measure sufficiently similar to GNP to appeal to mainstream economists, while offering substantial improvements by accounting for widely acknowledged negative developments associated with economic activity (Costanza et al., 2009; Giannetti et al., 2015). Notably, Daly and Cobb (1994, p. 379) acknowledge that "[t]he ISEW is far from perfect", viewing it as a temporary measure until more profound systemic changes are possible. Some of the literature discussed here also refers to the "Genuine Progress Indicator" (GPI), which is very similar to the ISEW and differs only in a few components (Cobb et al., 1995).

In the decades since its development, the ISEW has been replicated, refined, and applied widely. One strand of the literature focuses on improving its components. Following Daly and Cobb's (1994) own revision of the ISEW, several elements have been reassessed by numerous scholars. For example, ozone depletion was included in the original ISEW but later publications have argued for its removal (Bleys, 2008; Garcia, 2021; Lawn, 2003). Another long-standing debate concerns whether, and how, leisure time should be incorporated (Bagstad et al., 2014; Daly and Cobb, 1994; Lawn, 2003).

Other studies have calculated the ISEW for different geographical areas. Beyond the original assessment for the United States (Daly and Cobb, 1989), the ISEW has been calculated for all EU countries (Soupart and Bleys, 2024), several Asian countries (Clarke, 2004; Menegaki, 2018; Menegaki and Tugcu, 2018), many Global North countries (Kenny et al., 2019; Soupart and Bleys, 2024), and a few Latin American countries (Castañeda, 1999; Sánchez et al., 2020) and African countries (Menegaki and Tugcu, 2016). Sub-national assessments also exist, such as for California (Brown and Lazarus, 2018), Rio de Janeiro (Senna and Serra, 2021), regions in England (Jackson et al., 2008), and Chinese provinces (Zhu et al., 2022).

A third focus has been the ISEW's applications. Many studies examine its historical trajectory in comparison with GDP or other indicators to assess the desirability of past developments. In several cases, divergences between faster-growing GDP and more slowly growing or declining ISEW are attributed to environmental degradation accompanying economic growth (O'Mahony et al., 2018; Sánchez et al., 2020; Senna and Serra, 2021). The ISEW has also been used to test Max-Neef's (1995) threshold hypothesis, which posits that economic growth raises welfare only up to a point, after which further growth reduces it. However, findings on whether the hypothesis holds remain inconclusive (Daly, 1999; Tsara et al., 2024; Van der Slycken and Bleys, 2024). Similarly, the ISEW has been used to test the feasibility of decoupling economic welfare from resource use. Studies find that welfare as measured by the ISEW reacts more negatively to increased resource use than welfare measured by GDP (Beça and Santos, 2014; Kalimeris et al., 2020).

### 2.2 ISEW criticism and theoretical underpinning

Several scholars have criticised the ISEW, both with respect to its conceptual basis and methodological design. Conceptually, the ISEW has been identified as an indicator of weak sustainability, as its accounting framework allows the substitution of different environmental or social goals (Neumayer, 2025). Furthermore, the ISEW poses challenges for measuring progress in a post-growth economy: on the one hand, post-growth could increase the ISEW due to improvements in income equality and environmental preservation. On the other hand, a reduction in consumption would be counted negatively. It is therefore unclear whether the ISEW would rise or fall under a degrowth trajectory (O'Neill, 2012).

Methodologically, several components have been scrutinised and revised. However, these numerous modifications have raised concerns that the ISEW now consists of competing definitions, making meaningful comparisons between empirical applications difficult (Neumayer, 2025).

A further critique concerns the ISEW's lack of an explicit theoretical foundation. Lawn (2003) addresses this critique by arguing that the ISEW aligns with the Fisherian conceptualisation of income and capital, which measures the welfare directly experienced by individuals as a result of economic activity. More recently, Van der Slycken and Bleys (2020) have distinguished between an ISEW based on "Benefits and Costs Experienced" ($ISEW_{BCE}$) and one based on "Benefits and Costs of Present Activities" ($ISEW_{BCPA}$).

The $ISEW_{BCE}$ follows the Fisherian notion of income by measuring the welfare individuals experience from economic activity undertaken during the exact period and within the geographic area for which the indicator is calculated. The $ISEW_{BCPA}$, by contrast, goes beyond both geographic and temporal boundaries by accounting for benefits and costs that arise in the future or outside the geographic area where the measured activities occur.

More precisely, the $ISEW_{BCPA}$ largely relies on consumption-based accounting for environmental costs. This method quantifies a country's resource use and pollution based on environmental damages caused in the production, distribution, and use of goods and services consumed by its residents, rather than quantifying the resource use and pollution occurring within a country's territory (as in territorial or production-based accounting). The $ISEW_{BCPA}$ also accounts for a wider range of types of environmental degradation, including types that have no immediately perceivable effects but may cause future harm, such as nuclear energy generation or the depletion of non-renewable resources. In addition, the $ISEW_{BCPA}$ incorporates changes in the capital stock, which are excluded from the $ISEW_{BCE}$ because they have no immediate effect on experienced welfare (Van der Slycken and Bleys, 2023).

### 2.3 Ecological macroeconomic modelling

Since this study uses the ecological macroeconomic model COMPASS (Vogel et al., 2025), we briefly discuss the relevant literature. Ecological macroeconomic models have become a popular approach for analysing interdependence between the macroeconomy, society, and the environment (Hardt and O'Neill, 2017). Although these models can follow different modelling approaches, they share an acknowledgement of the economy's embeddedness within society and the biosphere, and of the profound effects these spheres exert on one another. These connections are often represented through a system dynamics approach, which captures dynamic interactions and nonlinear feedbacks between model components to understand complex systems (Van Eynde et al., 2024).

Rather than relying solely on aggregate mainstream indicators such as GDP, ecological macroeconomic models evaluate a diversity of outcomes (Van Eynde et al., 2024). They also typically reject several prominent assumptions of mainstream economic models, including optimal equilibria arising from perfectly rational agents, and the substitutability of factors of production (Hardt and O'Neill, 2017).

Many ecological macroeconomic models are designed to explore post-growth scenarios (Edwards et al., 2025; Lauer et al., 2025; Van Eynde et al., 2026). Post-growth serves as an umbrella term for approaches that question GDP growth as a policy objective, while prioritising environmental sustainability, human wellbeing, and social equity (Kallis et al., 2025; Van Eynde et al., 2026). Important post-growth models include EUROGREEN (D'Alessandro et al., 2020), LowGrow SFC (Jackson and Victor, 2020), and COMPASS (Vogel et al., 2025).

The COMPASS model is an ecological macroeconomic model specifically developed to evaluate policies against the "Doughnut" of social and planetary boundaries. The Doughnut, or safe and just space framework, is a post-growth approach intended to assess how societies perform in relation to social thresholds and ecological boundaries (Fanning and Raworth, 2025; Raworth, 2017). The COMPASS model of the Doughnut is currently parameterised for Spain (Vogel et al., 2025).

### 2.4 Social and environmental policies

Within this study, we integrate three different policies into the COMPASS model: a carbon tax, income redistribution, and working-time reduction. We selected these three policies because they collectively represent both mainstream and post-growth approaches and they affect different components of the ISEW. Below, we provide an overview of the rationale for each policy.

The first policy, a carbon tax, is widely supported in the mainstream literature for promoting environmental sustainability (Baranzini et al., 2017; Hepburn et al., 2020). The underlying idea is that a carbon tax raises the costs of emitting carbon, providing a market incentive for less carbon-intensive production and consumption. Its actual effect on emissions depends on several factors, including the local context, the tax rate, the treatment of traded goods, and the extent to which carbon-intensive goods and services can be substituted (Andersson and Atkinson, 2020; Köppl and Schratzenstaller, 2023; Wang et al., 2022).

The second policy, income redistribution, is a broad goal that can be implemented through various policy tools or their combination. Redistribution typically aims to achieve a more equal distribution of income, wealth, or both (Cowell, 2018; Kuypers et al., 2021). The policy goal is advocated by both pro-growth and post-growth researchers (Apostel and O'Neill, 2022; Kallis et al., 2025; Ostry et al., 2014; Stiglitz, 2016). Redistribution can generate a range of socio-economic effects, depending on context and policy design (Li et al., 2024).

The third policy, working-time reduction, is frequently proposed in the post-growth literature, particularly for the Global North. It has the potential to lower economic activity (Kallis, 2011; Nørgård, 2013), reduce unemployment (Jackson and Victor, 2020; Oberholzer, 2023; Vogel et al., 2024), and generate other beneficial social and environmental effects (Cieplinski et al., 2023, 2021). Proposals for working-time reduction differ in the role ascribed to labour productivity: many post-growth scholars propose reducing working time in proportion to the rise in hourly productivity, keeping aggregate output constant and converting productivity gains into increased leisure rather than increased output (Cieplinski et al., 2021; Dietz and O'Neill, 2013; Schor, 2025). Yet, working-time reduction can also be implemented independently of productivity, or with a reduction exceeding productivity gains. In this case, aggregate output is either reduced or maintained through hiring additional workers (D'Alessandro et al., 2020; Sekulova et al., 2013). Some proposals suggest compensating wages lost due to shorter hours to keep annual per-capita income from work constant (Gerold and Nocker, 2018; Schor et al., 2022), whereas others allow for income to be reduced (D'Alessandro et al., 2020; Sekulova et al., 2013).

### 2.5 Research gaps in modelling the ISEW

The literature on the ISEW to date has mainly focused on (1) methodological critiques and innovations, and (2) applications to historical periods in specific geographical areas. To the best of our knowledge, only one study has analysed possible future trajectories of the ISEW (Rugani et al., 2018). It uses an artificial neural network to forecast the ISEW for Luxembourg based on time-series data, comparing two energy-consumption scenarios with additional inputs for demographic developments and consumption expenditures. However, it does not use an ecological macroeconomic model, includes no feedback mechanisms between inputs, and applies machine learning only to estimate each factor's individual effect on the ISEW, rather than their combined effect.

Therefore, there remains a lack of research on simulations of the ISEW for future scenarios. In particular, no study has dynamically modelled the ISEW while considering feedback effects between its components and the wider socio-economic and environmental context. Moreover, analyses of the ISEW's response to a broader range of policies are still missing. Our study aims to address these important gaps.

## 3. Methods

To assess the effects of our three policies on the ISEW and to analyse its suitability as an indicator for the evaluation of social and environmental policies, we implement the ISEW in the COMPASS model and compare the ISEW's performance with other model indicators. Below we explain the COMPASS model (Section 3.1), describe our approach to modelling the ISEW and its components (Section 3.2), and present the policy scenarios (Section 3.3).

### 3.1 The COMPASS model

The COMPASS model (Vogel et al., 2025) is a system dynamics model designed to assess a diverse range of policies. It captures the interplay between economic, social, and environmental systems, and evaluate the effect of different policies on indicators in comparison to the social thresholds and biophysical boundaries included in the Doughnut framework (Fanning and Raworth, 2025; Raworth, 2017).

COMPASS consists of four modules: demographics, economy, sustainability, and wellbeing. The modules are interconnected, influencing one another simultaneously (see Figure 1).

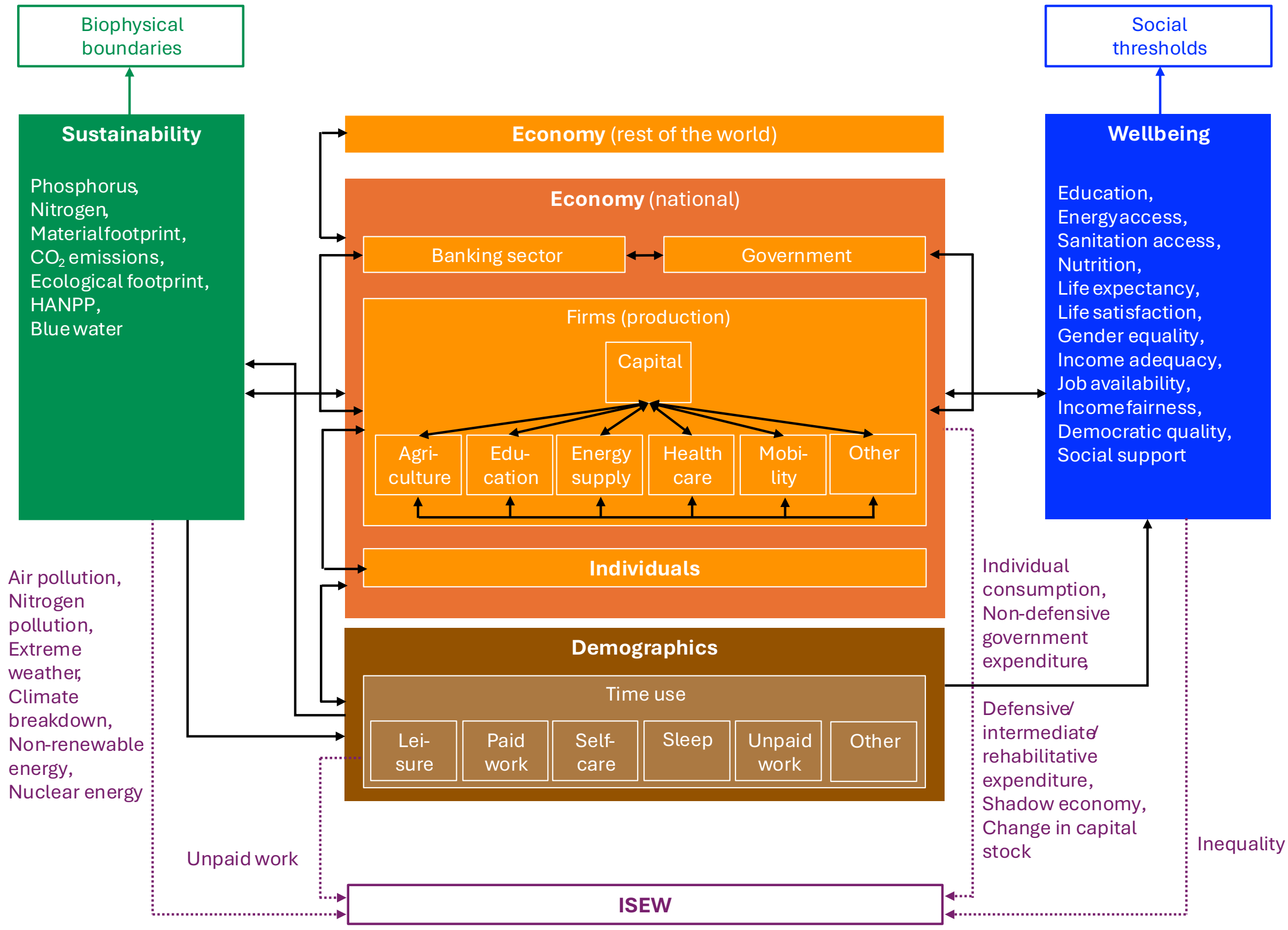

**Figure 1: Overview of the COMPASS model.** The model consists of four modules: economy, demographics, sustainability, and wellbeing. The elements shown in purple indicate the model variables used to calculate the ISEW. Source: adapted from Vogel et al. (2025).

The demographics module simulates population dynamics for different cohorts. The population is disaggregated by gender, age group, and skill level.

The economy module features an environmentally extended input–output structure based on the EXIOBASE database (Stadler et al., 2018) and is demand-driven. It comprises six sectors: agriculture, education, energy supply, health care, mobility, and other (which includes all remaining sectors). The model is stock–flow consistent and includes different behaviours for firms, individuals, the government, and the banking sector. Trade with the rest of the world is represented through exports and imports.

The sustainability module represents resource use, pollution, and Earth-system change. It is largely based on the environmental extension of the input–output structure with resource and pollution flows, enabling both production-based and consumption-based accounting. The module includes feedback effects of pollution on population health. It also features several environmental limits in terms of national biophysical boundaries, representing the equal-per-capita national allocation of global planetary boundaries featured in the Doughnut (Fanning et al., 2021; O'Neill et al., 2018; Vogel et al., 2025).

The wellbeing module calculates the Doughnut's social indicators. Each indicator has its own functional form and is driven by a combination of endogenous and exogenous variables. The social indicators are assessed against a social threshold, representing the minimum socially acceptable value (Fanning et al., 2021; O'Neill et al., 2018; Raworth, 2017; Vogel et al., 2025).

To assess whether the ISEW is well-suited for the evaluation of social and environmental policies, we draw on the model's capability to simulate GDP, environmental performance, and social outcomes, and compare the ISEW results with these indicators. A more detailed description of the model version used, including the operationalisation of biophysical boundaries and social thresholds, is provided by Vogel et al. (2025). In addition, the main modifications made to the COMPASS model for this study are described in the Supplementary Information (Annex I).

### 3.2 ISEW components

Our study models both the $ISEW_{BCE}$ and the $ISEW_{BCPA}$, following the definitions proposed by Soupart and Bleys (2024). The composition of each indicator is shown in Figure 2. The $ISEW_{BCE}$ includes the monetary evaluation of benefits from unpaid work, individual consumption expenditure, the shadow economy, and non-defensive collective government consumption. The costs comprise defensive, intermediate, and rehabilitative expenditure, a monetary equivalent of losses from income inequality, and a monetary evaluation of the damages of air pollution, nitrogen pollution, and extreme weather events.

The ISEW$_{BCPA}$ additionally accounts for monetarily evaluated costs stemming from climate breakdown, the depletion of non-renewable energy resources, and nuclear power use, while excluding extreme weather (assumed to be covered under climate-breakdown damages).[1] It also incorporates changes in the capital stock (investment minus depreciation).

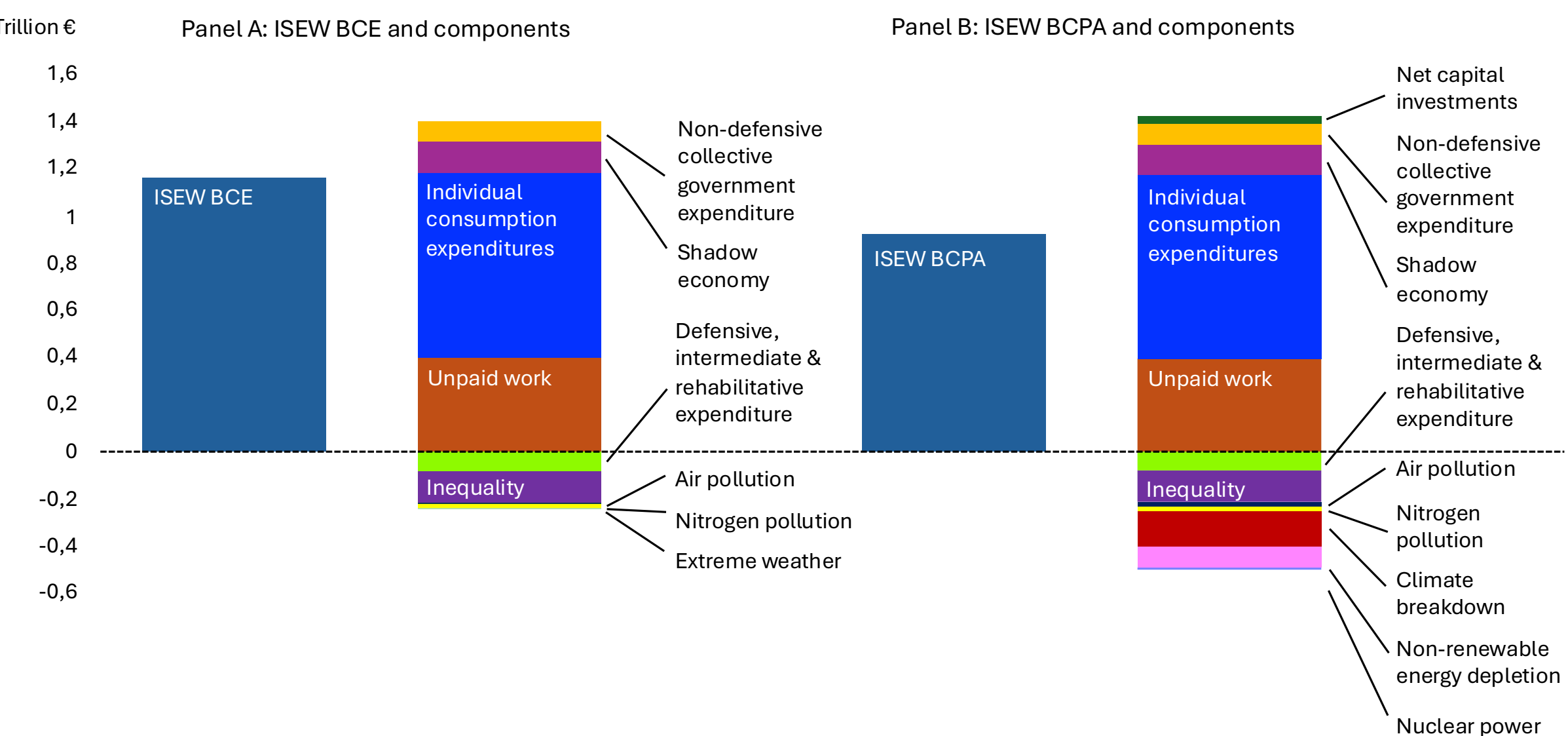

**Figure 2: Composition of the ISEW$_{BCE}$ and ISEW$_{BCPA}$.** Panel A presents the total value of the ISEW$_{BCE}$ for Spain in 2020, in trillion euros, on the left-hand side, and its individual components on the right. Panel B shows the corresponding values for the ISEW$_{BCPA}$. For both variants, positive values represent benefits, and negative values are deducted as costs.

We used a decision tree to include the ISEW components in COMPASS. Specifically, we applied three criteria: (1) whether the component or a suitable proxy was already part of the model, (2) whether the component had a relationship with existing model components, and (3) whether data were available to simulate it as a model variable (Figure 3).

[1] In Soupart and Bleys (2024), the environmental costs for the ISEW$_{BCE}$ are grouped as "narrow ecological costs", while those for the ISEW$_{BCPA}$ are grouped as "broad ecological costs". In addition to comprising different environmental pollutants and resources, the two categories also differ in their treatment of the same pollutant or resource: narrow ecological costs consider only immediate impacts, whereas broad ecological costs include long-term effects and impacts beyond national borders.

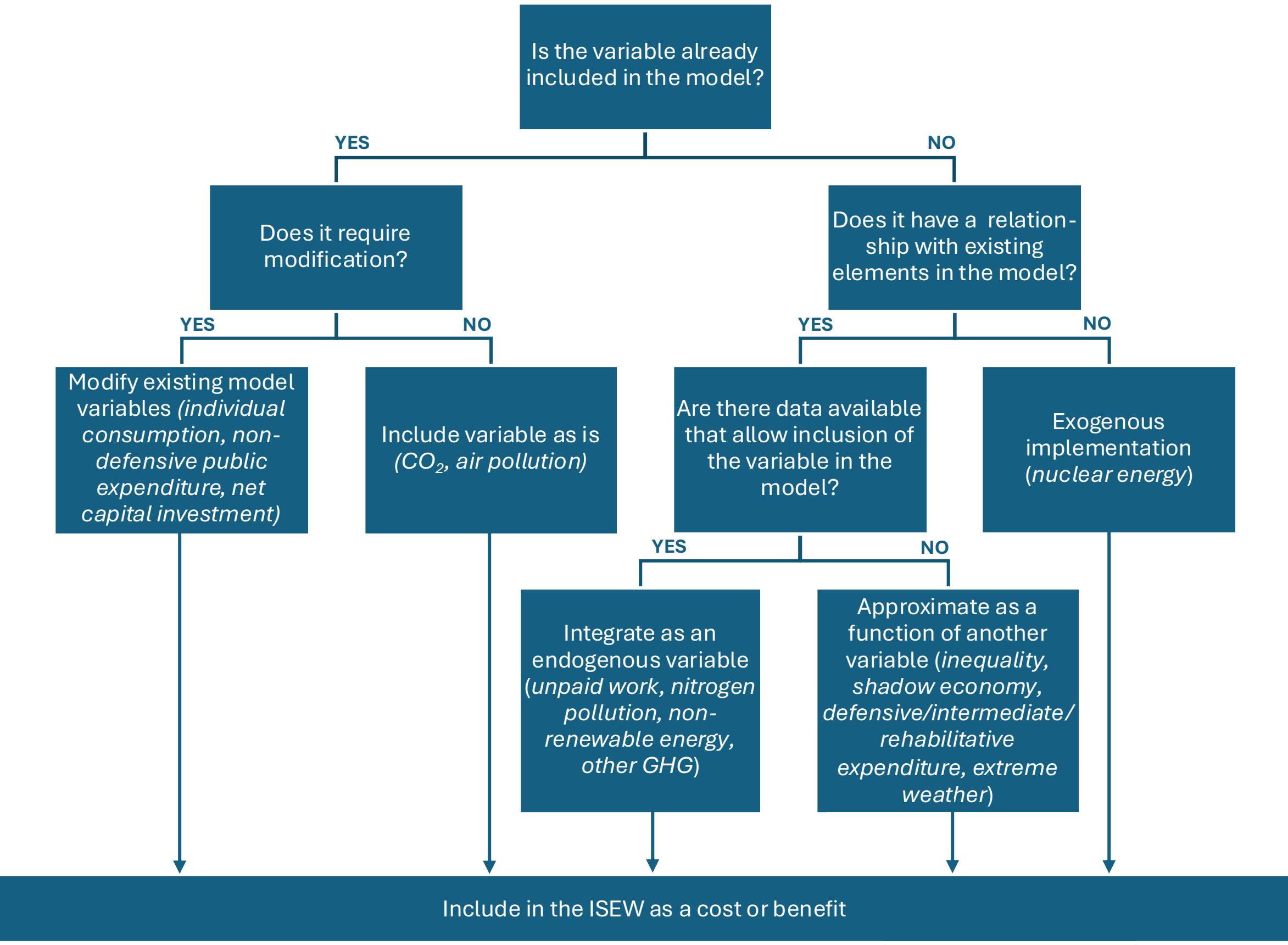

**Figure 3: Decision tree for integrating ISEW components into the COMPASS model.** Each box represents a decision point, starting at the top and proceeding downwards according to the answers. The italicised text lists the ISEW components implemented according to the strategy described in each box.

If none of these criteria were met, the component was integrated exogenously. If the component related to existing model features but data for endogenous simulation were unavailable, it was included as a share or factor of another endogenous variable, determined by the average historical value from 2000 to 2020. In such cases, historical values for the ISEW components were drawn from the $ISEW_{BCE}$ and $ISEW_{BCPA}$ data for Spain by Soupart and Bleys (2024). Their data were also used to validate whether the model implementation yielded realistic values.

Within the model, all components are combined to obtain the ISEW. Dividing this indicator by the Spanish population in the respective year (calculated in the demographics module) yields ISEW per capita. A detailed overview of the integration of the ISEW components is provided in the Supplementary Information (Annex I).

### 3.3 Modelling of policies

We implemented three policies (a carbon tax, income redistribution, and working-time reduction) in the COMPASS model. The policies are phased in over five years, starting in 2030 and reaching full effect in 2035, except for the carbon tax, which initiates a dynamic adjustment process in 2030. The simulation period runs from 2020 to 2070. A detailed description of all policies is provided in the Supplementary Information (Annex II).

The carbon tax is modelled as an endogenously determined tax rate on territorial emissions. The rate depends on the gap between actual and target emissions, a maximum $CO_2$ tax rate, and an adjustment-speed parameter. Based on a specified emissions-reduction rate associated with the maximum tax rate, the actual territorial $CO_2$ reduction is proportional to the applied rate. Assuming a border carbon adjustment mechanism, the same reduction rate is applied to consumption-based $CO_2$ emissions from trade. Thus, the policy primarily affects the ISEW by reducing $CO_2$ emissions. The macroeconomic effects are simplified, as the emission reduction follows an exogenously imposed relationship with the tax rate and is not linked to structural macroeconomic change.

Income redistribution is represented through a progressive taxation scheme and higher social benefits for some groups. From 2030, marginal tax rates are reduced for lower income groups and raised for higher income groups, stabilising in 2035 at 13% for the lowest tax bracket and 75% for the highest tax bracket (previously 19% and 47%, respectively). Since these are marginal rates, only the share of income within each bracket is taxed at the respective rate, while the remainder is taxed at rates for other brackets.

Social benefits are initially a fixed share of the average monetary wage of all employed people, with a uniform value for the entire population. The policy triggers a yearly increase in the social benefits for unemployed people and people out of the labour force, beginning in 2030 and stabilising in 2035. By the end of implementation, benefits have doubled for those out of the labour force and increased by 30% for unemployed people (who also receive unemployment benefits). Both benefits and tax rates are calibrated so that aggregate consumption remains similar to the level for business-as-usual, ensuring that the ISEW response is driven by changes in inequality rather than consumption.

Working-time reduction is implemented as a 15% reduction in working hours per worker in a standard week, phased in over five years. Given the high unemployment in Spain in 2020, we implement an economy-wide shortening of the working week beyond hourly productivity gains to observe effects on both employment and time use. To avoid the ISEW response being driven mainly by consumption, we combine the reduction with a proportional downscaling of per-capita wages. As the model is demand-driven, economy-wide working hours and aggregate output remain unchanged, meaning that additional workers are required to maintain production, thereby reducing unemployment.

This policy also affects time use. Following the approach used in the EUROGREEN model (Cieplinski et al., 2023), average time-use profiles for social groups (defined by employment status and gender) are generated, in our case using the Spanish time-use survey (Instituto Nacional de Estadística, 2025). Time use is grouped into paid work, unpaid work, sleep, selfcare, leisure, and other (which includes all unassigned time use categories). After policy activation, individual paid working hours fall by the same proportion as aggregate paid hours. The freed-up time is allocated to other categories in proportion to their initial shares of total time use. For the unpaid work component of the ISEW, yearly unpaid hours are summed across the population and multiplied by an equivalent market wage of €9.04, following Soupart and Bleys (2024), which is below the model's average hourly net wage of €12.28 in 2020.

In addition to the three individual policies, we implement an all-three-policies scenario in which all three policies are activated simultaneously, as well as a business-as-usual scenario with no policy activation.

## 4. Results

In what follows, we present the results of simulating the ISEW, GDP, and important social and environmental indicators for the different scenarios in COMPASS. First, we examine the trajectory of both ISEW versions under business-as-usual, the individual policies, and their combination (Section 4.1). We then contrast the ISEW with GDP for the all-three-policies and business-as-usual scenarios (Section 4.2). Finally, we present findings for selected environmental indicators (in comparison to biophysical boundaries) and social indicators (in comparison to social thresholds) under the different scenarios, enabling comparison of the ISEW with other indicators (Section 4.3).

### 4.1 The ISEW for different policy scenarios

For both ISEW variants, the all-three-policies scenario outperforms business-as-usual and all single-policy scenarios by the end of the simulation period (Figure 4). Per-capita values of the $ISEW_{BCE}$ and $ISEW_{BCPA}$ grow in every scenario, except for the $ISEW_{BCPA}$ in the working-time reduction and redistribution scenario, which plateau towards the end. While the $ISEW_{BCE}$ per capita increases almost linearly before and after policy implementation for all policies, the $ISEW_{BCPA}$ per capita grows at a declining rate after implementation for all policies.

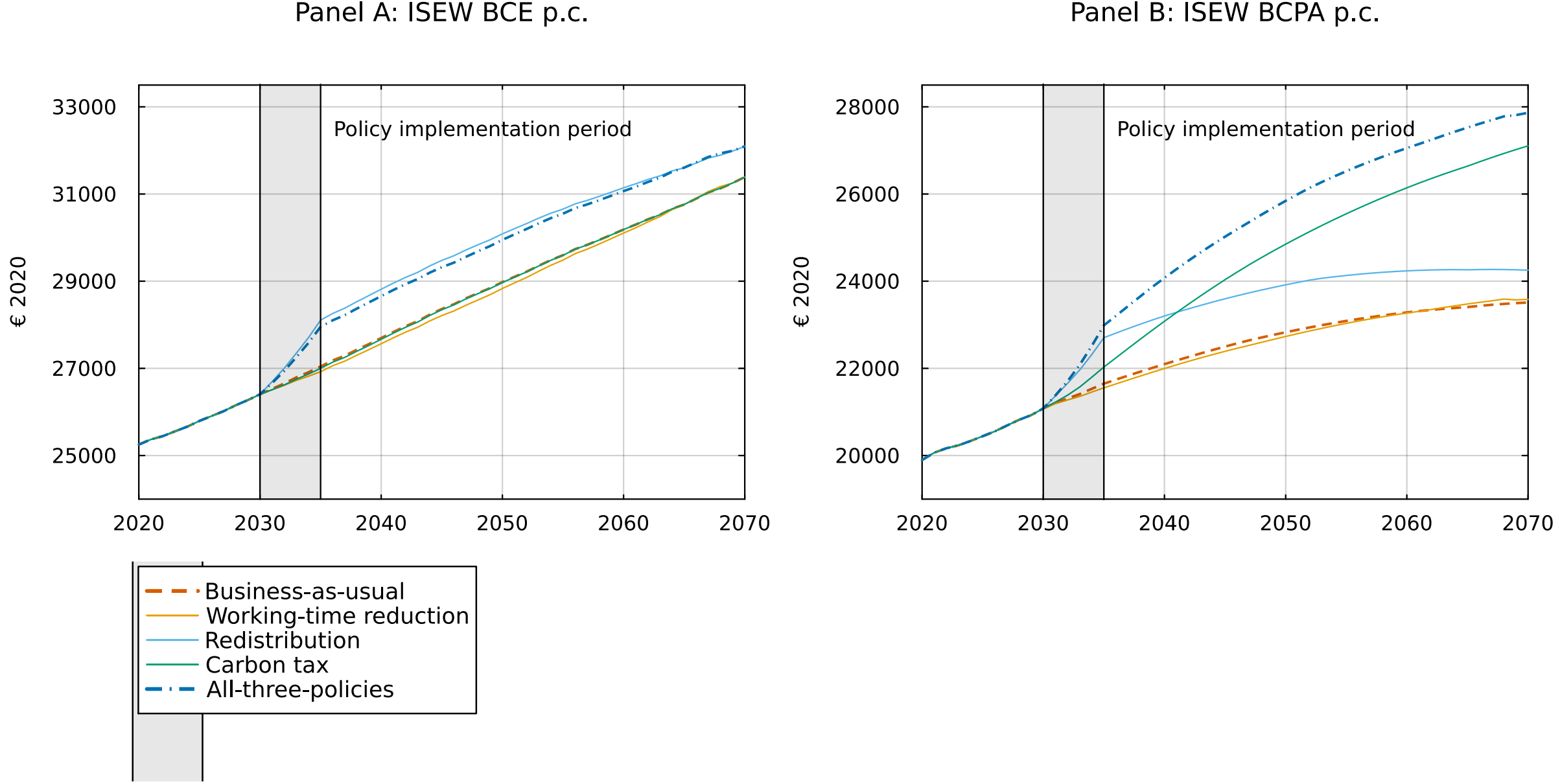

**Figure 4: Evolution of per-capita values of $ISEW_{BCE}$ and $ISEW_{BCPA}$ under different policy scenarios.** Panel A shows the evolution of $ISEW_{BCE}$ per capita in constant 2020 euros, and Panel B shows the evolution of $ISEW_{BCPA}$ per capita in constant 2020 euros. The lines represent the business-as-usual scenario, the all-three-policies scenario, and the three individual policy scenarios (carbon tax, redistribution, and working-time reduction). The period over which the policies are introduced is shown in grey.

For the $ISEW_{BCE}$, the all-three-policies scenario produces a higher growth rate than the business-as-usual scenario during the implementation period, after which the growth rates for these scenarios converge. For the $ISEW_{BCPA}$, however, the all-three-policies scenario leads to a higher growth rate than business-as-usual that is sustained after the implementation period. As a result, improvements in the $ISEW_{BCE}$ by the end of the simulation period are relatively small (around 2%), while improvements in the $ISEW_{BCPA}$ are substantial (around 18%).

For the individual policies, the carbon tax produces almost no difference compared to business-as-usual for the $ISEW_{BCE}$, since $CO_2$ emissions are not considered in this version and other impacts are minimal due to simplified macroeconomic dynamics. For the $ISEW_{BCPA}$, the carbon tax leads to a higher growth rate than business-as-usual because reduced $CO_2$ emissions lower the associated cost components. The trajectory runs parallel to the all-three-policies scenario, suggesting that the latter's growth rate is largely driven by the carbon tax.

In the redistribution scenario, both ISEW variants rise strongly during the policy implementation period, after which growth rates stabilise at levels similar to business-as-usual. Redistribution performs almost identically to the all-three-policies scenario for the BCE variant, but remains well below it for the BCPA variant. The initial rise in both versions stems from a 25% reduction in welfare losses from inequality over the implementation period. Over subsequent decades, inequality increases again across all scenarios, affecting absolute ISEW levels for all scenarios.

Lastly, working-time reduction does not outperform business-as-usual in either variant. Initially, the ISEW values are slightly lower than business-as-usual, and only by the end of the period does working-time reduction slightly exceed business-as-usual.

This counterintuitive result is driven by the dynamics of unpaid work. Employed individuals reduce paid hours, and part of the freed-up time is reallocated to unpaid work. However, reduced hours per capita lead to new hires, shifting some individuals from unemployment to employment. Because unemployed people devote a larger share of time to unpaid work, this shift reduces total unpaid hours at the societal level. Overall, unpaid work lost due to fewer unemployed people outweighs unpaid work gained by employed individuals, causing a temporary net reduction.

Over time, however, unpaid work becomes higher in the working-time reduction scenario than in business-as-usual, causing the ISEW trajectory to surpass business-as-usual. This reversal is driven by population dynamics: the number of unemployed people is calculated as the labour force minus the employed population, and the labour-force share declines with population ageing. Because unemployment is larger under business-as-usual, ageing reduces the number of unemployed people more substantially in that scenario. Since unemployed people contribute more unpaid hours per person than employed individuals, unpaid work declines more under business-as-usual, explaining the later crossover.

Because total paid work is redistributed between groups, working-time reduction has no direct effect on economy-wide paid hours and therefore no immediate effect on consumption. The model currently does not account for changes in the scale and composition of consumption by income group, which could occur in reality. Changes in other time-use categories (i.e. leisure) are not considered in our ISEW version. Consequently, the ISEW initially falls slightly after implementation of the working-time reduction.

### 4.2 Comparison of the ISEW and GDP

We now examine how the ISEW compares with GDP under the all-three-policies and business-as-usual scenarios. In contrast to both ISEW variants, GDP per capita shows only a minor reaction to the interventions (Figure 5). Its trajectory in the all-three-policies scenario (Figure 5b) is very similar to business-as-usual (Figure 5a). GDP per-capita growth is close to zero during the policy phase-in period (Figure 5b), after which it grows at a rate comparable to business-as-usual (Figure 5a).

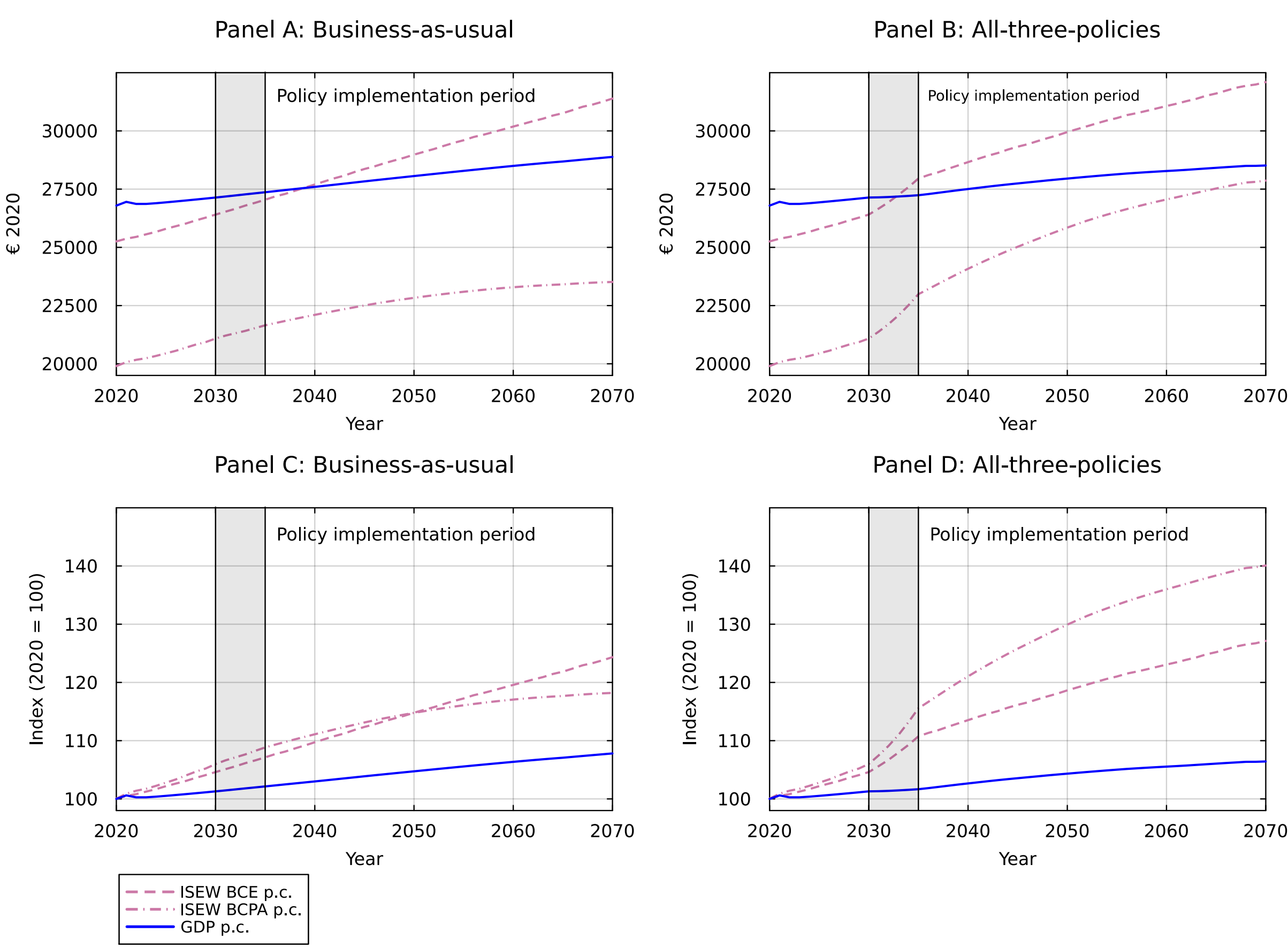

**Figure 5: Comparison of the ISEW to GDP under the all-three-policies and business-as-usual scenarios.** Panel A shows the evolution of the $ISEW_{BCE}$, $ISEW_{BCPA}$, and GDP in absolute per capita terms

under the business-as-usual scenario. Panel B shows the evolution of the $ISEW_{BCE}$, $ISEW_{BCPA}$, and GDP in absolute per capita terms under the all-three-policies scenario. Panel C shows the evolution of the $ISEW_{BCE}$, $ISEW_{BCPA}$, and GDP in per capita terms indexed to the base year under business-as-usual. Panel D shows the evolution of the $ISEW_{BCE}$, $ISEW_{BCPA}$, and GDP in per capita terms indexed to the base year under the all-three-policies scenario. The base year is 2020, with the index set to 100.

The baseline positive growth of both ISEW versions and GDP reflects rising labour productivity in the model, which increases production and consumption. This dynamic affects individual consumption expenditure, an important sub-component of both GDP and the ISEW (48% of the total absolute value of $ISEW_{BCE}$ components and 40% for $ISEW_{BCPA}$ in 2020).

Both ISEW variants lie below GDP at the start of the simulation (Figures 5a and 5b), indicating that the ISEW effectively incorporates some of the costs of harmful economic activity. The $ISEW_{BCPA}$ starts lower than the $ISEW_{BCE}$ because it accounts more extensively for environmental costs, considers a broader set of environmental components, and assigns higher costs due to footprint-based accounting, which yields higher values for pollution and resource use than territorial accounting in Spain.

In the business-as-usual scenario, the $ISEW_{BCPA}$ remains below GDP, while the $ISEW_{BCE}$ rises above GDP around 2038 (Figure 5a). In the all-three-policies scenario, the $ISEW_{BCE}$ rises above GDP during the policy phase-in period, and the $ISEW_{BCPA}$ nearly converges with GDP towards the end (Figure 5b). Although the ISEW penalises harmful activities, it also captures positive socio-economic changes stemming from policy interventions. These results suggest that the ISEW is more capable than GDP at reflecting the positive social and environmental effects of the policies (which we analyse further in Section 4.3).

Lastly, we evaluate normalised ISEW and GDP trajectories using 2020 as the base year. Both ISEW variants grow faster than GDP in the business-as-usual scenario (Figure 5c). Since GDP and the ISEW grow simultaneously, the negative effects of economic activity do not outweigh the benefits accounted for in the ISEW within our simulations. Only the $ISEW_{BCPA}$ shows a flattening trajectory under business-as-usual, indicating that costs approach benefits towards the end of the period. In the all-three-policies scenario, both ISEW variants grow steeply during the implementation period due to positive policy effects (Figure 5d). These findings suggest that the negative effects of economic activity play a relatively minor role in the ISEW's evolution.

### 4.3 Social thresholds and biophysical boundaries for different policy scenarios

Finally, we assess how the five policy scenarios affect selected social indicators in comparison to thresholds associated with meeting basic needs, and environmental indicators in comparison to downscaled planetary boundaries (Figure 6). Social indicators should ideally rise above their thresholds, while environmental indicators should remain below their boundaries.

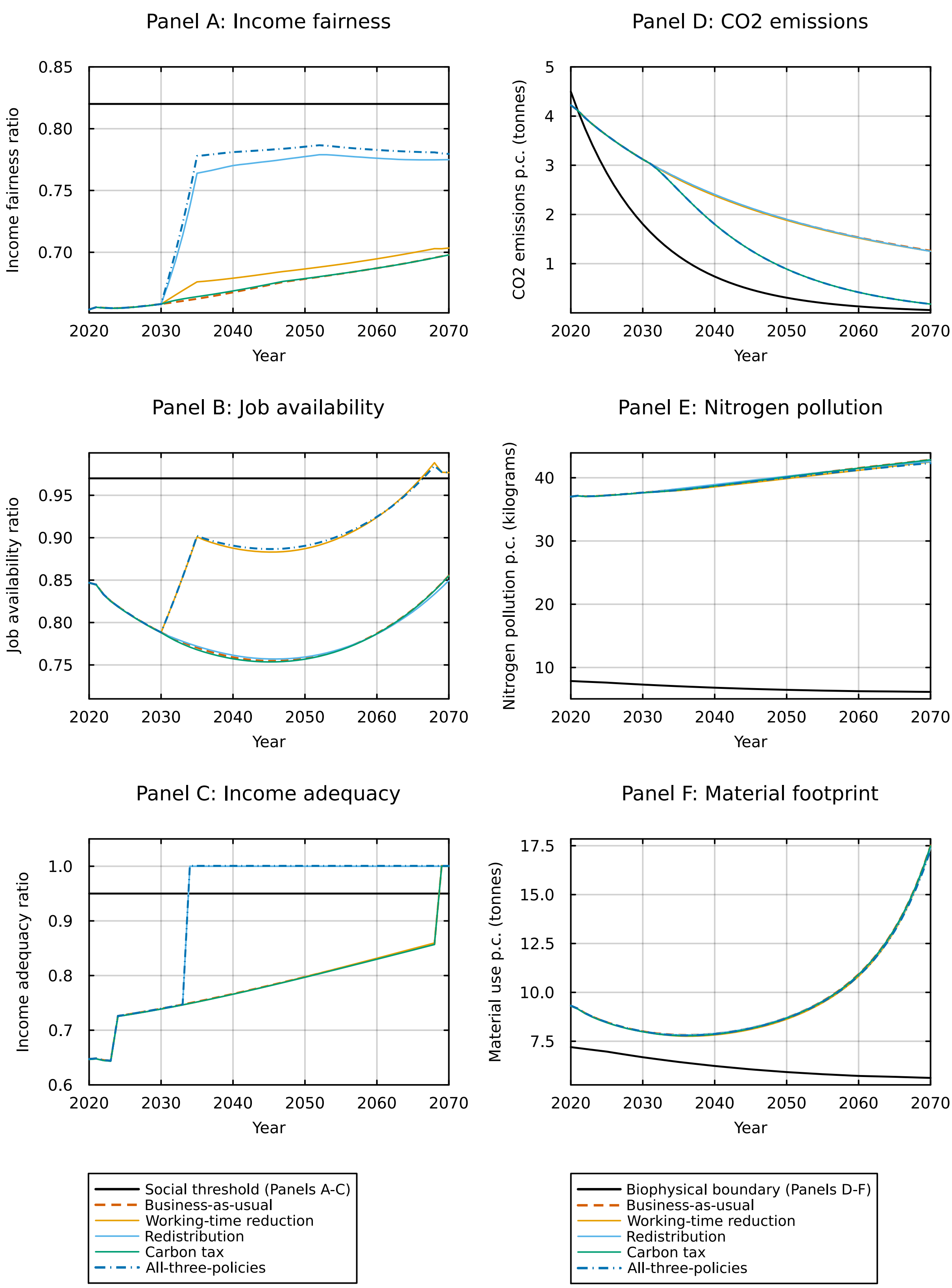

**Figure 6: Effects of the different policy scenarios on selected social and environmental indicators, in comparison to their respective thresholds and boundaries.** Panels A-C show social indicators, while Panels D-F show environmental indicators. The coloured lines represent the outcomes of the business-as-usual scenario, the all-three-policies scenario, and the three individual policy scenarios (carbon tax, redistribution, and working-time reduction). The solid black lines represent the social thresholds and

biophysical boundaries from the Doughnut of social and planetary boundaries. Ideally, the social indicators would rise above the social thresholds, while the environmental indicators would remain below the biophysical boundaries.

Of the social indicators calculated by COMPASS, we consider income fairness (a measure of equality), job availability (a measure of employment), and income adequacy (the inverse of poverty), as they show the strongest policy response. Income fairness performs best under the all-three-policies scenario, showing substantial improvement under redistribution and a slight improvement under working-time reduction relative to business-as-usual (Figure 6a).

Job availability responds strongly to working-time reduction during implementation. Afterwards, it follows a pattern of initial decline and later growth across scenarios. As job availability is the inverse of the unemployment rate, the initial decline after implementation is driven by rising labour productivity, which reduces labour demand. The subsequent reversal is driven by a shrinking labour force, implying that the number of unemployed people — the residual of the labour force minus employed people — also shrinks relative to employed people. The threshold for job availability (i.e. full employment) is met for working-time reduction and the all-three-policies scenario by the end of the simulation, while the other three scenarios remain below it (Figure 6b).

Income adequacy improves strongly under redistribution and the all-three-policies scenario during the policy implementation period, reaching the threshold associated with eliminating poverty (Figure 6c). For the other three scenarios, it is reached shortly before the end of the period. The sharp jumps in the trajectory of income adequacy are explained by its measurement as the share of the population with an income above the poverty line. Given the low resolution of income groups in the model, income adequacy changes markedly when a group crosses the line. The continuous growth of income adequacy under business-as-usual is explained by ongoing shifts in population group sizes. Overall, improved income fairness and income adequacy stem from redistribution, while increased job availability and, to a lesser extent, improved income adequacy are caused by working-time reduction, indicating that both policies achieve the expected positive impacts.

Regarding environmental indicators, we analyse $CO_2$ emissions, nitrogen pollution, and material footprint, as these are both represented in COMPASS and included in the ISEW's environmental costs by Soupart and Bleys (2024)[2]. For $CO_2$ emissions, although the biophysical boundary is transgressed in all scenarios, the degree of overshoot is substantially lower under the carbon tax scenario and all-three-policies scenario, reflecting the positive impact of the carbon tax (Figure 6d).

[2] The material footprint is not directly included in the ISEW components proposed by Soupart and Bleys (2024). However, it substantially overlaps with the ISEW component of non-renewable energy depletion, as the material footprint in the COMPASS model accounts for the domestic extraction of fossil fuels, crops, and other biomass (next to metal ores and non-metallic minerals) (Vogel et al., 2025). Disregarding the category of nuclear energy, the non-renewable energy used and produced in Spain consists of fossil fuels, biofuels, and waste products (International Energy Agency, 2026).

For nitrogen pollution and material footprint, all scenarios follow a similar trajectory. Nitrogen pollution grows throughout (Figure 6e), while material footprint grows after the policy-implementation period (Figure 6f). Material footprint follows a steeper growth trajectory than nitrogen pollution because their development depends on the respective intensity of output as well as the growth rate of the output intensity, which is calculated based on the trend of past values (see Vogel et al., 2025). While nitrogen pollution has a negative output-intensity growth rate and therefore grows less strongly despite increasing output, the material footprint has a positive intensity growth rate and thus grows faster. Both indicators lie far above their limits at the end of the simulation, indicating substantial environmental damage.

Overall, the biophysical indicators present a less optimistic picture of environmental and social sustainability in Spain than the ISEW suggests. Despite substantial improvements in social outcomes, the results show that the policy interventions are insufficient to address the severe overshoot of biophysical boundaries.

## 5. Discussion

Overall, the all-three-policies scenario outperforms business-as-usual and single-policy scenarios for both ISEW variants. The strongest positive single-policy effects arise from redistribution for the $ISEW_{BCE}$ and from the carbon tax for the $ISEW_{BCPA}$. Of the policies we tested, only working-time reduction initially yields a lower ISEW than business-as-usual because of a net reduction in unpaid work. Comparing the ISEW with GDP shows that the ISEW is more responsive to policy effects than GDP. Both ISEW variants grow throughout the simulation period due to rising productivity, which translates into rising consumption. The ISEW in our simulation assigns a positive weight to consumption growth, while the negative consequences of consumption have only a limited impact on its trajectory. While the ISEW shows that our three policies have positive impacts relative to business-as-usual, it does not show what the individual social and environmental indicators reveal — a severe overshoot of biophysical boundaries at the end of the simulation.

We now discuss the results, beginning with the study's limitations (Section 5.1). We then examine how our study contributes to the literature (Section 5.2), discuss the value of dynamically modelling the ISEW (Section 5.3), and assess the suitability of the ISEW for evaluating social and environmental policies (Section 5.4).

### 5.1 Limitations of the study

Our analysis has several limitations that should be acknowledged. First, given the complexity of the socio-economic and environmental systems simulated by COMPASS, the model has certain shortcomings. The model currently assumes a fixed propensity to consume and a fixed consumption composition across socio-economic groups. Outside of carbon tax, demand is unresponsive to price signals, as rising prices trigger inflation but not behavioural change. Furthermore, the model does not endogenously represent technological change or structural change in provisioning systems. These limitations primarily affect the ISEW consumption component, which links to several other components. Moreover, household granularity is limited to 30 broad socio-economic groups, constraining measurement of inequality and income adequacy. The labour-force participation rate is constant, adding rigidity to employment dynamics and thus to the benefits from unpaid work in the ISEW. Regarding the environment, a more complete set of feedback mechanisms onto society, and more detailed modelling of environmental impacts from changes in provisioning systems, could substantially alter the ISEW's environmental costs. While Vogel et al. (2025) highlight the lack of consumption-based accounting as a limitation, it has been incorporated in this study, as explained in the Supplementary Material (Annex I).

Second, while policies are modelled as described in Section 3, they could also have been implemented differently. For the carbon tax, the link between the tax and $CO_2$ reduction is exogenously quantified, while further impacts on the economic system are not represented. Achieving the emissions reductions implied by the policy would likely require profound changes in consumption and production (Köppl and Schratzenstaller, 2023), including altered consumption patterns (Chen et al., 2025), firm behaviour (Gallé et al., 2025), and employment (Azevedo et al., 2023), which could affect the ISEW. Including these dynamics would require a more detailed macroeconomic representation of the carbon tax. However, in our study, the implementation of the carbon tax is focused on observing its effect on the ISEW via $CO_2$ emissions.

Since redistribution is more of a goal than a specific policy, different measures could also have been applied. Notably, wealth redistribution is neglected in our study although wealth inequality is high in Global North countries (Skopek et al., 2014; Zucman, 2019) and strongly affects social mobility and wellbeing (Sałach-Dróżdż, 2024; Schechtl, 2025). Yet, our study focuses on income redistribution, as this is the most widely discussed dimension of economic inequality (Nolan et al., 2011).

For working-time reduction, two main alternatives exist: a version where hours are reduced in line with rising productivity (Cieplinski et al., 2021) and a version with a compensation of wages lost to shorter hours (Hanbury et al., 2023; Schor et al., 2022). We chose a configuration independent of productivity and without wage compensation to observe the impacts of employment, time use, and inequality on the ISEW, without them being overshadowed by consumption effects.

Third, because we model some components as dependent on individual consumption expenditure, our results likely exaggerate the correlation between the ISEW and GDP. Most importantly, the defensive, intermediate, and rehabilitative expenditure category is modelled as a fixed share of individual consumption, and the shadow economy as a fixed share of GDP. Notwithstanding this limitation, when summing up both benefit and cost components of the ISEW in absolute values, defensive, intermediate, and rehabilitative expenditure is only 5% of the $ISEW_{BCE}$ and 4% of the $ISEW_{BCPA}$ in 2020, and the shadow economy is 8% of the $ISEW_{BCE}$ and 7% of the $ISEW_{BCPA}$. Other major components like unpaid work (24% of $ISEW_{BCE}$), inequality (8% of $ISEW_{BCE}$), and government expenditure (5% of $ISEW_{BCE}$) are modelled independently of individual consumption. Hence, our modelling approach slightly exaggerates the ISEW's dependency on consumption. However, the high positive weight given to the component of actual individual consumption in the ISEW is an inherent feature of the index's methodology.

### 5.2 Contributions of this study to the literature

Notwithstanding these limitations, our study makes three important contributions. First, we incorporate the ISEW into a dynamic simulation model for the first time, allowing us to derive insights regarding future ISEW trajectories. We are able to show how the ISEW evolves when interactions between its components are considered simultaneously, rather than estimating the impact of individual factors based on historical relationships, as previous work has done (e.g. Rugani et al., 2018).

Second, our study is the first to test the effect of three important policies (a carbon tax, redistribution, and working-time reduction) on the ISEW. The three assessed policies mostly have a positive effect, and the best outcome arises when all three policies are combined. We find that redistribution has the largest positive impact on the $ISEW_{BCE}$ and a carbon tax has the largest positive impact on the $ISEW_{BCPA}$. In other words, identifying the most favourable policy depends on which ISEW version is used.

Finally, our study makes an important conceptual contribution by comparing the ISEW's ability to capture policy effects with that of other indicators, in particular those linked to the Doughnut of social and planetary boundaries. The comparison allows us to better assess the ISEW's suitability for evaluating social and environmental policies.

### 5.3 The value of dynamically modelling the ISEW

Our results show the value of dynamically modelling the ISEW, as they reveal developments that are otherwise difficult to anticipate, given the multitude of factors that simultaneously influence the ISEW. The same intervention may push different components in opposite directions or generate counteracting effects within the same component. Dynamic modelling allows us to capture the net effect of these interactions.

Surprisingly, we find that the ISEW decreases slightly in the working-time reduction scenario due to a decline in unpaid hours contributed by unemployed individuals. Although unpaid work increases among employees and welfare losses from inequality fall substantially, these effects are insufficient to generate a positive net outcome. Dynamic modelling helps to show the outcome of these complex interactions.

Our approach also sheds light on the ISEW's response to economic growth under the different scenarios, which is difficult to predict because growth in GDP increases both benefits and costs included in the indicator (O'Neill, 2012). While our results are bound to the features of the system represented in our model, we find that GDP and the ISEW increase together in all tested scenarios. Hence, for the simulated scenarios, penalties for environmental damages from economic growth are outweighed by benefits attributed to other components.

Furthermore, our study helps to understand differing responses across the two ISEW variants. While the $ISEW_{BCPA}$ suggests that a carbon tax has the strongest positive impact (Figure 4b), the $ISEW_{BCE}$ shows no reaction to the policy (Figure 4a). Moreover, the all-three-policies package only slightly alters the growth trajectory of the $ISEW_{BCE}$, whereas it shifts the $ISEW_{BCPA}$ from an almost flat trajectory to one that far outpaces the $ISEW_{BCE}$ (Figures 5b and 5c).

5.4 Suitability of the ISEW for the evaluation of social and environmental policies

Our results show that both ISEW versions respond more strongly to social and environmental policies than GDP does, as illustrated by the steep increase during the policy implementation period (Figure 5). Thus, the ISEW is arguably superior to GDP in evaluating socially and environmentally desirable outcomes.

However, the ISEW is also strongly driven by individual consumption expenditure, which comprises 48% of the $ISEW_{BCE}$ and 40% of the $ISEW_{BCPA}$. In addition, individual consumption indirectly affects other components, such as the shadow economy, defensive, intermediate, and rehabilitative expenditure, and several environmental costs (nitrogen pollution, air pollution, greenhouse gas emissions, and primary energy use). The latter are modelled as time-varying intensities of production and consumption (see Annex I). Not only does the ISEW tend to increase with GDP, it also assigns a substantial negative weight to decreasing consumption. Hence, the ISEW may be ill-suited to measuring profound socio-economic changes, such as those required for a post-growth economy, an issue previously raised by O'Neill (2012).

When compared with selected social thresholds and biophysical boundaries, a key strength of the ISEW is that it summarises trade-offs between components in a single measure. This one-dimensionality makes the indicator intuitive to use and, in principle, provides a clear picture of whether a scenario improves or worsens over time (Figure 4) (Jackson, 2004). Yet this ISEW property entails a substantial drawback. By collapsing all components into one value, the ISEW obscures individual trajectories, allowing improvements in some areas to mask deterioration in others (Neumayer, 1999; Ziegler, 2007). This substitutability underscores the ISEW's design as an indicator of weak sustainability rather than strong sustainability (Dietz and Neumayer, 2007; Neumayer, 1999) and becomes problematic when it conceals degradation in components crucial to planetary stability or human wellbeing.

Furthermore, the ISEW only enables relative assessment across years or places. It does not indicate whether a socio-economic pathway is sufficient to meet standards of planetary sustainability or high human wellbeing (Brennan, 2008; Dietz and Neumayer, 2007; Raworth, 2017). For example, the $ISEW_{BCPA}$ reacts positively to $CO_2$ reductions from a carbon tax (Figure 4), but evaluation against biophysical boundaries reveals that emissions still exceed the sustainable level (Figure 6). Empirical evidence shows that even ecosystems formally designated for protection can experience substantial human appropriation of net primary production (HANPP), highlighting how relative improvements can mask absolute ecological overshoot (O'Neill and Abson, 2009).

The comparison with environmental sustainability indicators also highlights the selectiveness of the ISEW's components. While some indicators linked to planetary boundaries are represented within the ISEW (specifically $CO_2$ emissions and nitrogen pollution), the majority are omitted (biodiversity loss, land-system change, phosphorus pollution, water use, ocean acidification, atmospheric aerosol loading, ozone depletion, and novel entities). Previous research has strengthened the theoretical underpinnings of the accounting method of the ISEW (Lawn, 2003; Van der Slycken and Bleys, 2020), but a robust framework specifying which environmental and social components to include is still lacking (Neumayer, 1999).

Finally, the ISEW's core idea of netting benefits against costs requires translating developments in different spheres into a single unit. Because this unit is monetary value, the approach entails philosophically and ethically contentious operations (Ekins et al., 2003; Neumayer, 1999; Ziegler, 2007), such as assigning market values to critical environmental resources or human life years (used to assess pollution damage) (Soupart and Bleys, 2024).

## **6. Conclusion**

This study is the first to dynamically model the ISEW, addressing a clear gap in the literature. We integrate the ISEW into the COMPASS ecological macroeconomic model, and investigate two questions: (1) How do three different policies, namely a carbon tax, redistribution, and working-time reduction, affect the ISEW? (2) Is the ISEW a suitable indicator for evaluating social and environmental policies?

Regarding the first question, our results show that the ISEW per capita increases under almost all policy scenarios, as well as business-as-usual. The best results for the ISEW occur when all three policies are combined. The most effective single policy is income redistribution when assessed with the $ISEW_{BCE}$, and a carbon tax when assessed with the $ISEW_{BCPA}$. Of the policies we tested, only working-time reduction generates an initial decline below business-as-usual, suggesting that it is not beneficial from the ISEW perspective.

Regarding the second question, GDP increases throughout the simulation but is almost unaffected by the policies. Thus, our results suggest that the ISEW is superior to GDP for evaluating social and environmental policies, as it shows a more pronounced policy response than GDP. However, the ISEW is unable to show whether the policies can achieve environmental or social goals in absolute terms, and its trajectory is strongly driven by consumption expenditure, similar to GDP. For guiding societies towards sustainable and inclusive wellbeing, a multi-dimensional approach that includes social thresholds and biophysical boundaries, such as the Doughnut framework, may be preferable.

**Declaration of generative AI and AI-assisted technologies in the manuscript preparation process**

During the preparation of this article the authors used Microsoft Copilot to improve the clarity of the language. After using this tool, the authors reviewed and edited the content and take full responsibility for the content of the published article.

**Acknowledgements**

The authors thank Claire Soupart for her helpful feedback which improved this work.

**Funding**

This work was supported by the European Union in the framework of the Horizon Europe Research and Innovation Programme under grant agreement numbers 101094211 (ToBe: “Towards a Sustainable Wellbeing Economy”), and 101137914 (MAPS: “Models, Assessment, and Policies for Sustainability”).